\documentclass{article}

\usepackage{arxiv}
\usepackage[utf8]{inputenc} 
\usepackage[T1]{fontenc}    
\usepackage{hyperref}       
\usepackage{url}            
\usepackage{booktabs}       
\usepackage{amsfonts}       
\usepackage{nicefrac}
\usepackage{microtype}
\usepackage{lipsum}	
\usepackage{graphicx}
\usepackage{doi}
\usepackage{subcaption}

\usepackage{fancyhdr} 
\usepackage{graphicx}
\usepackage{booktabs}
\usepackage{amsmath}
\usepackage{multirow}
\usepackage{makecell}

\usepackage{amsmath,amssymb,amsfonts}
\usepackage{algorithmic}
\usepackage{graphicx}
\usepackage{textcomp}
\usepackage{xcolor}
\usepackage[numbers]{natbib}
\usepackage{hyperref}
\usepackage{makecell}
\usepackage{textcomp}
\usepackage{array}
\usepackage{float}
\usepackage{stfloats}
\renewcommand{\toprule}{\hline}
\renewcommand{\midrule}{\hline}
\renewcommand{\bottomrule}{\hline}
\usepackage{geometry}

\title{Model Cascading for Code: A Cascaded Black-Box Multi-Model Framework for Cost-Efficient Code Completion with Self-Testing}

\date{} 					

\author{
    Boyuan Chen\textsuperscript{1,2}, 
    Mingzhi Zhu\textsuperscript{1}, 
    Brendan Dolan-Gavitt\textsuperscript{1}, 
    Muhammad Shafique\textsuperscript{2}, 
    Siddharth Garg\textsuperscript{1} \\
    \textsuperscript{1}Tandon School of Engineering, New York University, NY, USA \\
    \textsuperscript{2}eBRAIN Lab, Division of Engineering, New York University Abu Dhabi, UAE \\
    \texttt{\{boyuan.chen, mz3379, brendandg, muhammad.shafique, sg175\}@nyu.edu}
}

\renewcommand{\headeright}{} 
\renewcommand{\undertitle}{}
\renewcommand{\shorttitle}{}

\hypersetup{
pdftitle={Model Cascading for Code: Reducing Inference Costs with Model Cascading for LLM Based Code Generation},
pdfsubject={cs.AI, cs.LG},
pdfauthor={Author Names},
pdfkeywords={Model Cascading, LLM, Code Generation, Inference Optimization},
}

\begin{document}
\maketitle

\begin{abstract}
The rapid advancement of large language models (LLMs) has significantly improved code completion tasks, yet the trade-off between accuracy and computational cost remains a critical challenge. 
While using larger models and incorporating inference-time self-testing algorithms can significantly improve output accuracy, they incur substantial computational expenses at the same time.
Furthermore, servers in real-world scenarios usually have a dynamic preference on the cost-accuracy tradeoff, depending on the budget, bandwidth, the concurrent user volume, and users' sensitivity to wrong answers.
In this work, we introduce a novel framework combining model cascading and inference-time self-feedback algorithms to find \emph{multiple near-optimal} self-testing options on the cost-accuracy tradeoff in LLM-based code generation. 
Our approach leverages self-generated tests to both enhance accuracy and evaluate model cascading decisions. As a blackbox inference-time method, it requires no access to internal model parameters. We further propose a threshold-based algorithm to determine when to deploy larger models and a heuristic to optimize the number of solutions, test cases, and test lines generated per model, based on budget constraints.
Experimental results show that our cascading approach reduces costs by an average of \emph{26\%}, and up to \emph{70\%} in the best case, across various model families and datasets, while maintaining or improving accuracy in natural language generation tasks compared to both random and optimal single-model self-testing schemes.
To our knowledge, this is the first work to provide a series of choices for optimizing the cost-accuracy trade-off in LLM code generation with self-testing.
\end{abstract}

\section{Introduction}
\begin{figure*}[b]
    \newlength{\hhavg}
    \newlength{\wwavg}
    \newlength{\ttavg}
    \newlength{\movebackavg}
    \newlength{\legendhhavg}
    \setlength{\hhavg}{5.3cm}
    \setlength{\wwavg}{3.2cm}
    \setlength{\movebackavg}{-0.18cm}
    \setlength{\legendhhavg}{5.9cm}
    \setlength{\ttavg}{0.6cm}

    \includegraphics[width=\hhavg, trim={\the\wwavg{} 0 \the\wwavg{} \the\ttavg{}}, clip]{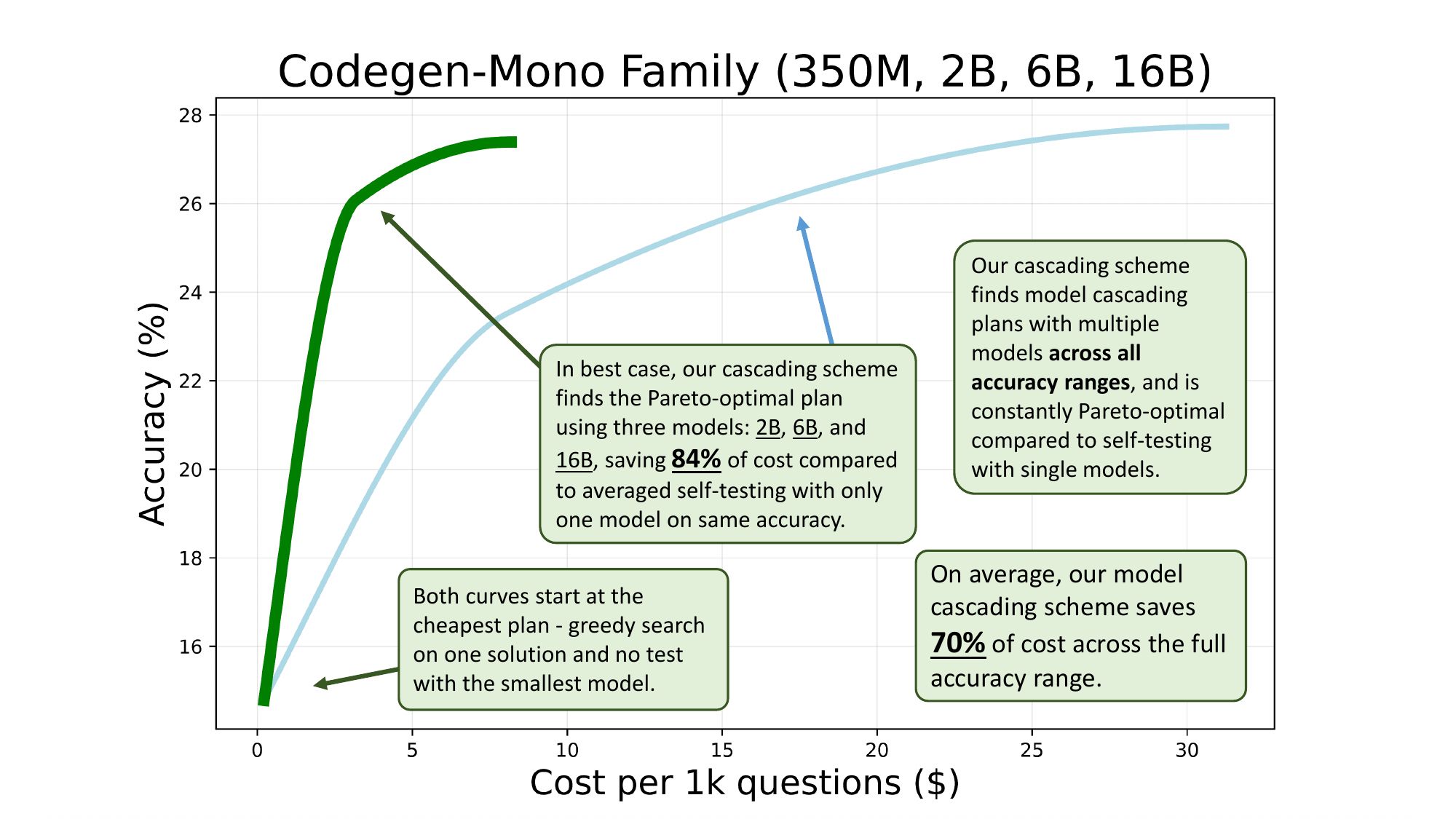}
    \includegraphics[width=\hhavg, trim={\the\wwavg{} 0 \the\wwavg{} \the\ttavg{}}, clip]{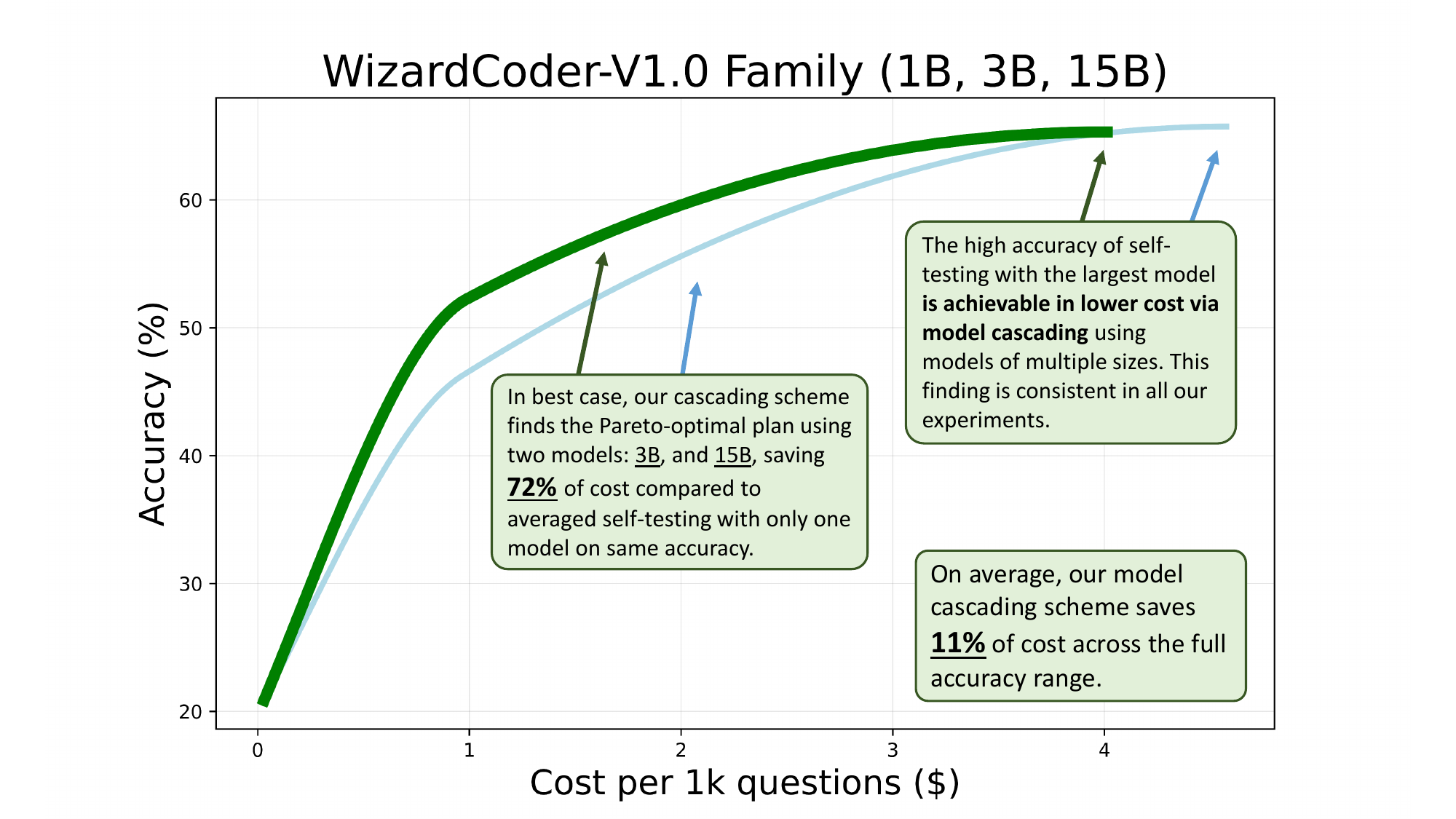}
    \includegraphics[width=\hhavg, trim={\the\wwavg{} 0 \the\wwavg{} \the\ttavg{}}, clip]{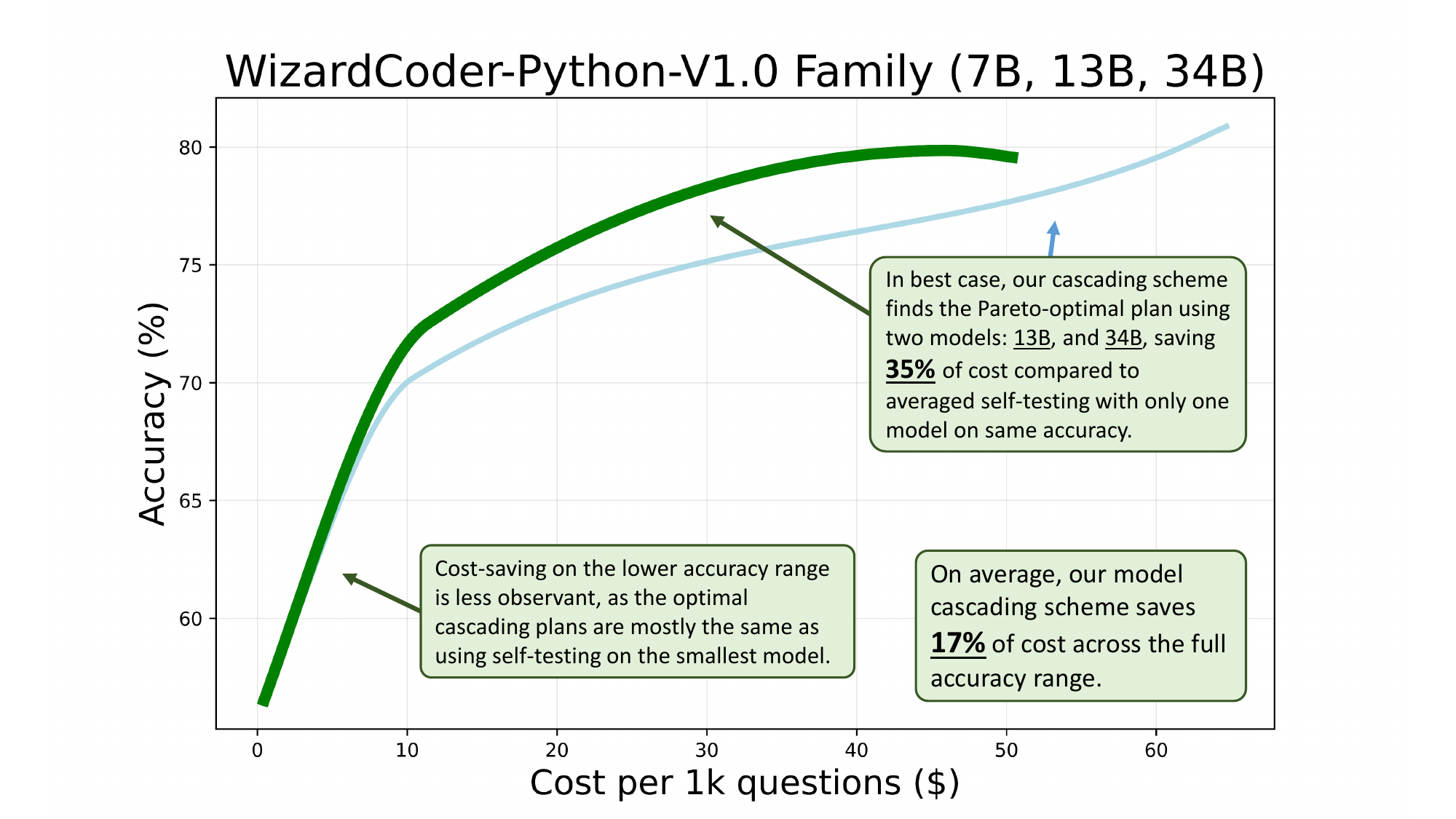}
    \includegraphics[width=\textwidth, trim={0 1.0cm 0 0.45cm}, clip]{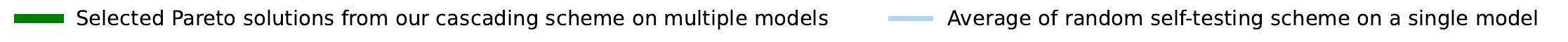}
    \vspace{-0.6cm}
    \caption{Results of our model cascading scheme compared to the randomly selected single-model self-testing scheme at a given accuracy on three model families solving the HumanEval dataset. The curves are derived via PCHIP interpolation from test set results in different parameter combinations, each represented as a dot in the cost-accuracy plot. On average, our scheme saves 70\%, 11\%, and 17\% of cost across all accuracy ranges on each model family. Detailed computation is provided in Section \ref{sec:compare_avg}.}
    \label{fig:avg_humaneval}
\end{figure*}

Large language models (LLMs) trained on extensive code datasets have demonstrated remarkable proficiency in programming tasks~\citep{guo2024deepseekcoderlargelanguagemodel, rozière2024codellamaopenfoundation}. 
Given the users' pursuit of pass@1 accuracy, researchers have developed a self-testing scheme, where the LLMs generate multiple answers and tests, and then return the answer that achieves the highest score in the tests \cite{tong-zhang-2024-codejudge, chen2022codet}.
This method is in parallel with the general approach in improving LLMs' output quality with self-generated feedback, as LLMs tend to generate more accurate verifications than answers \cite{NEURIPS2023_91edff07, huang2024selfimprovementlanguagemodelssharpening}.
While existing methods achieve significant accuracy improvements, they neglected the fact that the computational cost of LLMs in generating a series of code completions and tests remains prohibitively high \cite{zhu2023survey}, requiring \emph{multiple} forward passes of \emph{large} models on \emph{multiple} prompts. The uniqueness of user prompts further prevents caching, exacerbating computational demands. As a result, reducing inference costs while maintaining high accuracy is a critical challenge for scaling the self-testing scheme in LLM code completion systems.

\begin{figure}[H]
    \centering
    \begin{minipage}{0.55\textwidth}
        \centering
        \captionsetup{type=table}
        \renewcommand{\arraystretch}{1.2}
\centering
\small
\newlength{\topspace}
\newlength{\bottomspace}
\setlength{\topspace}{0.25em}
\setlength{\bottomspace}{0.12em}
\begin{tabular}{>{\centering\arraybackslash}p{2.1cm} 
                >{\centering\arraybackslash}p{0.6cm} 
                >{\centering\arraybackslash}p{1.0cm} 
                >{\centering\arraybackslash}p{0.8cm} 
                >{\centering\arraybackslash}p{0.8cm} 
                >{\centering\arraybackslash}p{0.8cm} 
                }
\toprule
\addlinespace[0.32em]
\multicolumn{2}{c}{LLM Family} & \multicolumn{1}{c}{Cost (\$)} & \multicolumn{3}{c}{Accuracy (\%)} \\
\cmidrule(lr){1-2} \cmidrule(lr){3-3} \cmidrule(lr){4-6}
Name & Size &  & H.Eval & MBPP  & APPS \\
\addlinespace[\bottomspace]
\midrule
\addlinespace[\topspace]
\multirow{3}{*}{Wizard-Py-V1.0} & 7B   & 2.22 & 56.7 & 53.9 & 16.4 \\
                                    & 13B  & 7.87 & 64.6 & 55.0 & 21.4 \\
                                    & 34B  & 23.34 & 72.6 & 61.8 & 23.7 \\
\addlinespace[\bottomspace]
\hline
\addlinespace[\topspace]
\multirow{3}{*}{Wizard-V1.0}        & 1B   & 0.23 & 23.8 & 33.0 & 3.9 \\
                                    & 3B   & 0.45 & 34.8 & 41.5 & 8.6 \\
                                    & 15B  & 3.22 & 60.4 & 53.2 & 13.8 \\
\addlinespace[\bottomspace]
\hline
\addlinespace[\topspace]
\multirow{4}{*}{Codegen-Mono}       & 350M & 1.28 & 14.6 & 22.0 & 0.4 \\
                                    & 2B   & 3.12 & 26.2 & 34.2 & 4.9 \\
                                    & 6B   & 7.96 & 27.4 & 41.9 & 5.1 \\
                                    & 16B  & 17.40 & 30.5 & 45.4 & 5.9 \\
\addlinespace[\bottomspace]
\bottomrule
\end{tabular}
\vspace{0.2cm}
        \caption{Cost and pass@1 accuracy with greedy search of all models on each dataset. Cost is estimated as dollars per one million tokens (\$/1M tokens) averaged on the three datasets. See details of cost calculation in Section \ref{sec:cost_calculation}. The APPS dataset refers to the introductory questions in the test set only.}
        \label{table:model_dataset_cost_accuracy}
    \end{minipage}%
    \hfill
    \begin{minipage}{0.4\textwidth}
        \centering
        \includegraphics[width=\textwidth]{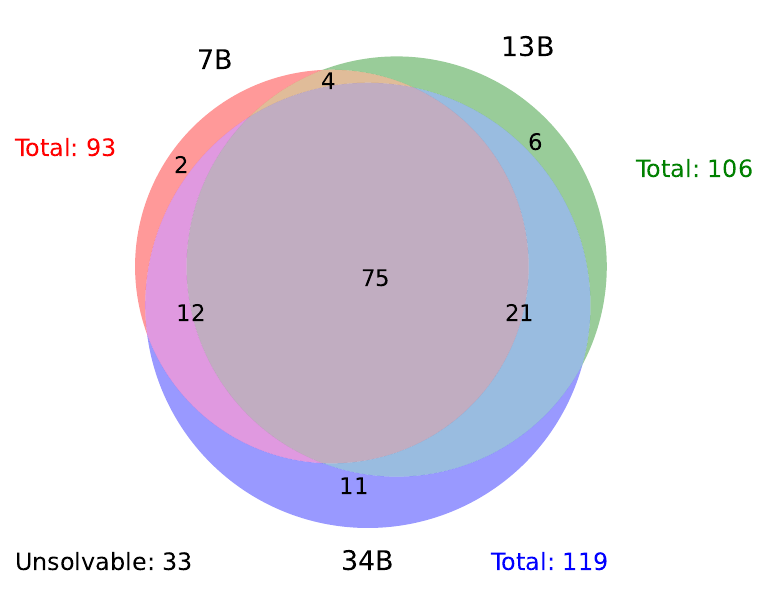}
        \caption{\small Venn diagram of WizardCoder-Python-V1.0 models (7B, 13B, 34B) on answering HumanEval prompts using greedy search. Out of 164 prompts, numbers indicate questions correctly answered by one, two, or all three models. Examples: 21 by 13B and 34B, 2 only by 7B, 75 by all. 33 questions were unsolved by any model.}
        \label{fig:venn_diagram}
    \end{minipage}
\end{figure}

To tackle this challenge, we introduce \textbf{a cascaded multi-model framework with self-testing}, which equips servers with a range of heuristically optimized plans across varying cost budgets. This enables dynamic adaptation to diverse user requirements and device constraints. Our approach leverages model families to strike a balance between accuracy and cost. For example, the smallest 7B model of Wizard-Python-V1.0 \cite{luo2023wizardcoder} achieves 56.7\% accuracy on HumanEval, while the largest 34B model reaches 72.6\% accuracy at approximately 10 times the cost (see Table \ref{table:model_dataset_cost_accuracy}). The Venn diagram in Figure \ref{fig:venn_diagram} further illustrates that the sets of solvable questions for each model size significantly overlap—over 90\% of the questions solved by the 34B model can also be solved by one of the smaller models. Interestingly, some questions are answered correctly only by smaller models within the family. By cascading from smaller to larger models and using self-generated tests to determine when to escalate to larger models, our framework identifies a set of optimal solutions that achieve higher accuracy at a lower cost compared to relying on a single model.

\begin{figure}[t]
    \centering
    \includegraphics[width=\textwidth, trim=0cm 0.9cm 0cm 0cm]{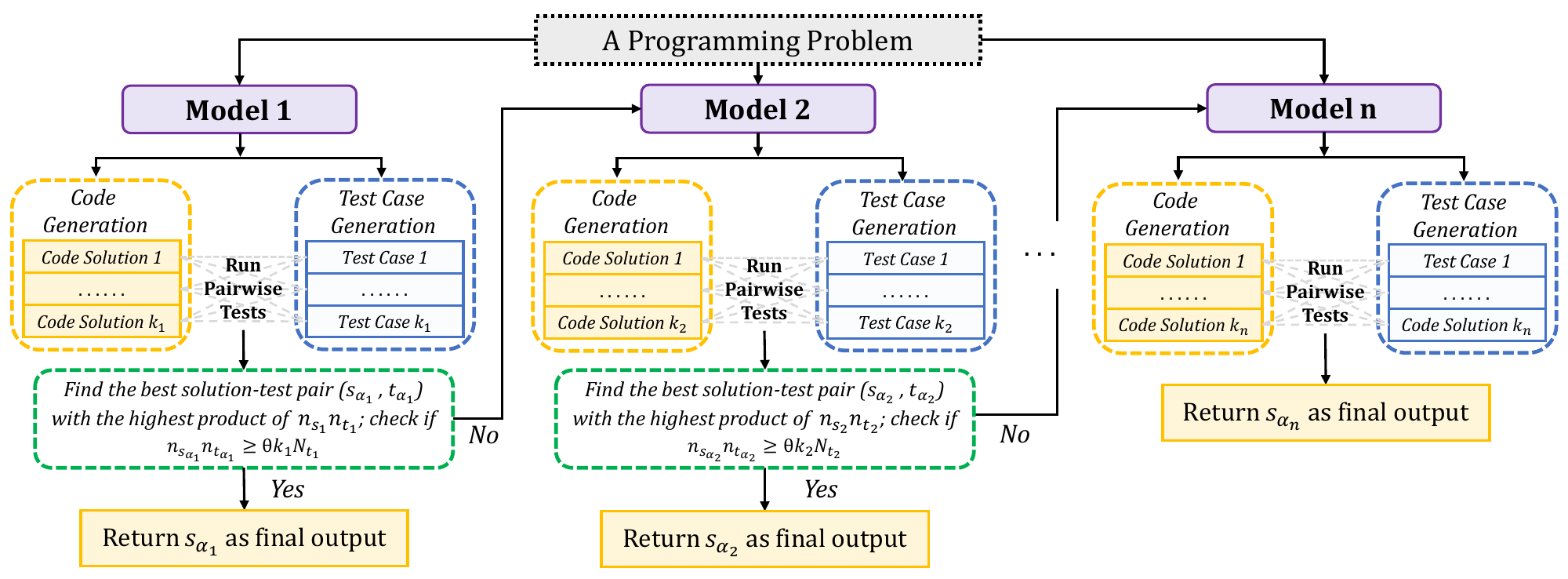}
    \vspace{0.1cm}
    \caption{An overview of our proposed model cascading solution with $n$ models. Models with higher indices are larger in size. Starting with model 1, we generate multiple code solutions and testcases. We score each solution-test pair and identify the best pair; if the score exceeds a threshold, we accept the solution and output; otherwise, we move on to the next model and repeat the process, until we take the highest-scored output from the largest model.}
    \label{fig:pipeline}
\end{figure}

Existing methods to reduce LLM inference costs, such as quantization~\cite{lin2024duquant}, pruning~\cite{qiao-etal-2024-threshold}, knowledge distillation~\cite{luo-etal-2024-amr}, and altered attention mechanisms~\cite{lv2024scalable}, focus on optimizing inference on a single model, under-exploiting the fact that models of various sizes are available. While speculative decoding~\cite{leviathan2023fast} can largely reduce inference time, it takes more memory to run both the target and the draft models and only performs outstandingly on small batches. On the other hand, model cascading~\cite{yue2024large} stands out for its ability to adapt to varying hardware constraints, batch sizes and user preferences, making it an ideal candidate for integration with self-testing schemes. 
Our framework leverages these insights, enabling models to generate both code solutions and test cases, with a learned threshold determining when to escalate to a larger model.

Our cascaded multi-model framework generates up to $k$ potential solutions and a set of test cases for each prompt. A learned threshold for the success rate determines when to seek assistance from a larger model, ensuring that the cascade stops once the pass rate exceeds the threshold. The optimal combinations of model selection, the number of answers and tests to generate, and the threshold parameter are learned from a pre-selected validation set that matches the test set's difficulty level. These parameters are identified by locating Pareto-optimal combinations on the accuracy-cost graph, and our experiments demonstrate that the validation set's Pareto combinations closely align with those of the test set.

To the best of our knowledge, ours is the \emph{first} solution to leverage a family of \emph{black-box} LLMs to provide \emph{multiple} near-optimal solutions for the tradeoff between operational cost and accuracy. While we evaluate model families differing in parameter size, our approach is easily extensible to other methods of generating model families, such as varying quantization levels. Furthermore, we reason that our method is compatible with speculative decoding, as it can also be included in the considered combinations.

Our empirical results show that the optimal cascading strategy robustly identifies cost-efficient and accurate generation plans. In the best case, our model cascading scheme reduces computation cost by \emph{70\%} while maintaining the same accuracy as using a single model with self-testing.

The contributions of this work are summarized as follows:
\begin{itemize}
    \item We propose a model-cascading framework for optimizing code generation with self-testing across multiple models. Our approach balances inference cost and pass@1 accuracy across all budgets by selecting \emph{Pareto-optimal} points from the cost-accuracy trade-off curve of the validation set.
    \item We empirically demonstrate that our model-cascading approach consistently achieves lower costs than randomly selected self-testing plans using a single model while maintaining the same level of accuracy. In the best case, our method reduces costs by up to \emph{70\%}.
    \item Our approach is designed for industrial code completion servers, is compatible with any multi-LLM system, and operates entirely in a black-box manner.
\end{itemize}

\section{Related Works}
\label{chapter:related_work}
\noindent \textbf{LLMs for Code} There is a large and growing body of work on LLM-based code generation. Recent fine-tuned LLM models for code include 
Qwen-Coder\cite{hui2024qwen25codertechnicalreport}, 
DeepSeek Coder \cite{guo2024deepseekcoderlargelanguagemodel}, 
Code Llama \cite{rozière2024codellamaopenfoundation}, etc. These models are typically evaluated using Pass@\textit{k} metrics. That is, they generate $k$ solutions and ``pass" a test as long as at least one solution is correct. In practice, however, a developer would have to pick from the $k$ solutions, requiring significant human effort.

\noindent \textbf{LLM Cascading} Prior work has investigated cascading for natural language prompts. FrugalGPT \cite{chen2023frugalgpt} trains a small DistillBERT model \cite{sanh2020distilbert} to output a score based on the query and the answer, and uses this score to determine when to query a larger model. 
So far, there is no published work in model cascading for code completion tasks, in part because output evaluation is more challenging.

\noindent \textbf{Self Testing for Code Generation} 
To assist models in ranking their own solutions, CodeT \cite{chen2022codet} has proposed that LLM models, in addition to solutions, generate test-cases as extra outputs, and run solutions through these test-cases to rank the best solutions. In our work, we utilize their solution-test score for cascading threshold. While CodeT's goal is only to rank-order solutions from a single model, model cascading on the other hand uses the same self-testing scheme to determine when to query the next larger model in the casade.

\section{Methodology}
\begin{figure*}[t!]
    \hspace{-0.33cm}
    \includegraphics[width=17cm, trim=0 0.7cm 0 0]{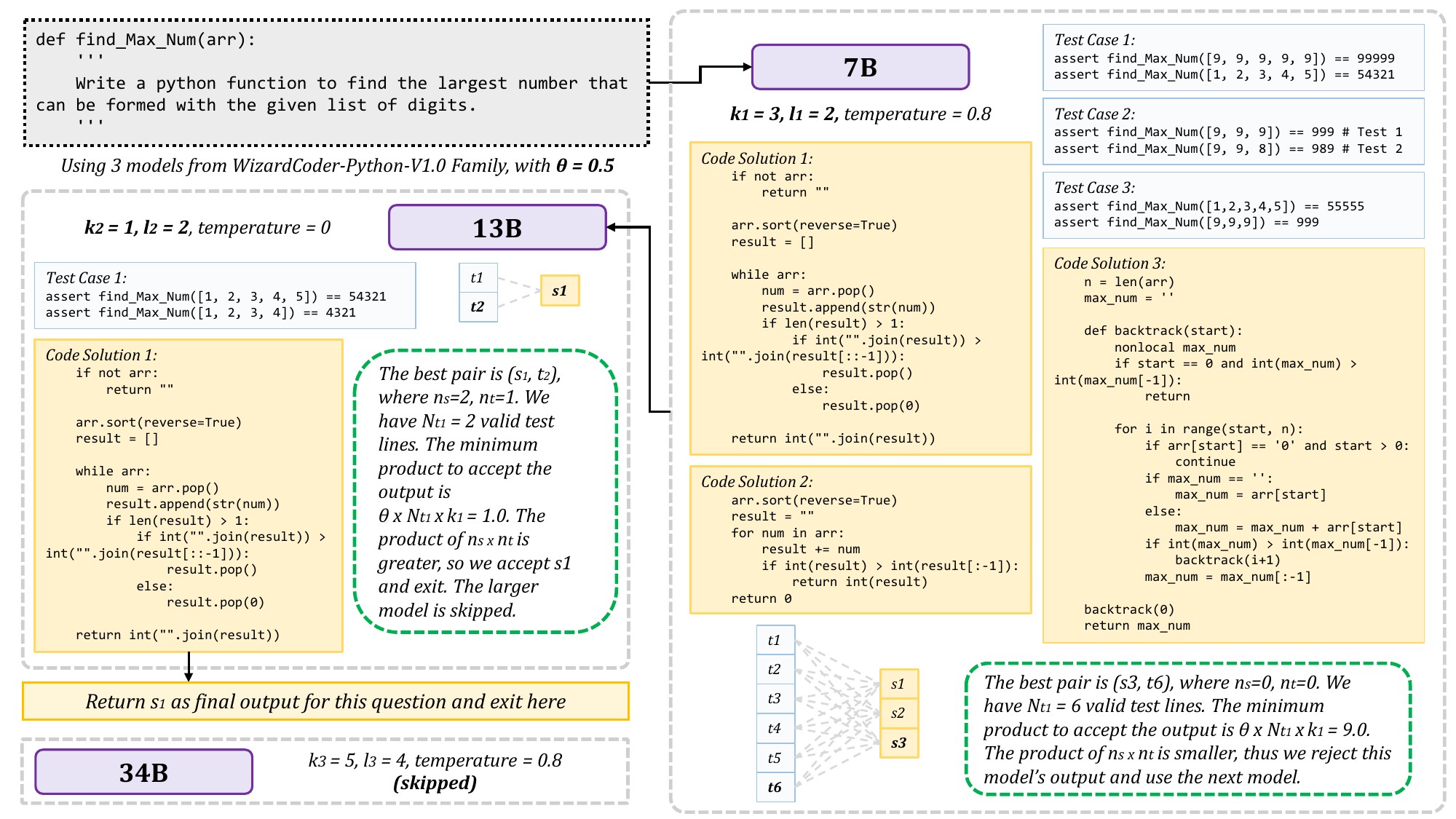}
    \vspace{-0.1cm}
    \caption{Solving MBPP question 57 "Find Max Num" with the WizardCoder-Python-V1.0 family of three model sizes: 7B, 13B and 34B. The threshold parameter in this case is threshold parameter $\theta=0.5$. The question is first passed to the 7B model, which generates $k_1=3$ solutions and test cases, each test case including $l_1=2$ test lines. There are $3 \times 2 \times 3=18$ solution-test pairs in total, and 0 pair passes, which is below the threshold $0.5 \times 18 = 9$. We thus pass the question to the next model in line, 13B. It generates $k_2=1$ solutions and test cases, each test case including $l_2=2$ test lines. There are $1 \times 1 \times 2 = 2$ solution-test pairs in total, and 2 of them pass, which is above the threshold. We therefore take the solution and exit. The biggest model with 34B parameter is skipped. }
    \label{fig:sampled_pipeline}
\end{figure*}
\vspace{-0.2cm}
\vspace{0.2cm}
We describe our model cascading approach for code generation, starting with how queries escalate to larger models. Figure \ref{fig:pipeline} provides a graphical overview, and Figure \ref{fig:sampled_pipeline} illustrates an example.

\subsection{Cascading Pipeline}
\label{subsect:cascading_pipeline}
Consider a model family $\mathcal{M}$ comprising $n$ models: $\{m_1, \ldots, m_n\}$, ordered by parameter size from smallest (and fastest) to largest. For a user prompt $p$, a natural language description of the desired function, we first query the smallest model, $m_1$, to generate $k_1$ solutions and $k_1$ corresponding tests. Each test consists of up to $l_1$ test lines, specifically \emph{assert} statements, as illustrated in Figure \ref{fig:pipeline}. The test lines from all solutions are pooled into a test suite $T_1$, which contains up to $k_1 \times l_1$ test lines. This suite is used to evaluate each solution.


Following prior work in self-testing \cite{chen2022codet, xiong2023program}, we adopt a similar strategy for selecting the best solution from multiple candidates. For each solution $s \in \mathcal{S} = \{s_1, \ldots, s_{k_1}\}$, we define $T_{pass} \subseteq T_1$ as the subset of test lines that $s$ passes. We then compute two quality metrics for $s$:
\begin{itemize}
    \item $n_s = |T_{pass}|$, the number of passing test lines; and
    \item $n_t = \max_{t \in T_{pass}}(solves(t,\mathcal{S}))$, the maximum number of solutions that pass any single test line in $T_{pass}$.
\end{itemize}
Here, $solves(t,\mathcal{S})$ denotes the number of solutions in $\mathcal{S}$ that pass test line $t$. Intuitively, $n_t$ reflects the quality of the passing test lines, with higher values indicating that a test line is non-trivial and robust, as it is passed by multiple solutions. The overall solution quality is quantified as the product $n_s \times n_t$, which we use to rank solutions. The highest-scoring solution-test pair is denoted ($s_{\alpha_1}$, $t_{\alpha_1}$), with a score of $n_{s_{\alpha_1}} \times n_{t_{\alpha_1}}$.


To determine whether a solution is acceptable, we introduce a threshold parameter $\theta \in [0, 1]$. Let $N_{t_1}$ denote the total number of valid test lines in $T_1$, where $N_{t_1} \leq k_1 \times l_1$ (since the model may stop generating before reaching the maximum number of test lines). If the highest solution score $n_{s_{\alpha_1}} \times n_{t_{\alpha_1}}$ satisfies:
\[
n_{s_{\alpha_1}} \times n_{t_{\alpha_1}} \geq \theta \times k_1 \times N_{t_1},
\]
we adopt the solution-test pair ($s_{\alpha_1}$, $t_{\alpha_1}$) and terminate the pipeline. Otherwise, we escalate the query to the next larger model, $m_2$, repeating the process: generating $k_2$ solutions and $k_2$ tests with up to $l_2$ test lines each, and scoring the solutions. This procedure continues until a solution exceeds the threshold $\theta$. When the largest model, $m_n$, is reached, we bypass the threshold check and directly adopt the highest-scoring solution.


\subsection{Finding Optimal Parameters}
\label{subsect:finding_parameters}
This section outlines the procedure for selecting optimal parameters in our cascading system. The system involves three primary sets of parameters: 
\begin{itemize}
    \item The number of solutions and tests generated for each model, denoted as $k_1, k_2, \dots, k_n$;
    \item The number of test lines in each test for each model, denoted as $l_1, l_2, \dots, l_n$; and
    \item A global cascading threshold parameter, $\theta$.
\end{itemize}

\paragraph{Parameter Ranges and Constraints}
We begin by defining the ranges and constraints for each parameter. For the number of solutions and tests ($k$), we consider the set $k \in \{-1, 0, 1, 3, 5, 10\}$ for each model. The value of $k$ determines the behavior of the system:
\begin{itemize}
    \item $k = -1$: The model is skipped.
    \item $k = 0$: One solution is generated using greedy search, and the system exits without generating tests.
    \item $k = 1$: One solution and one test are generated using greedy search.
    \item $k > 1$: $k$ solutions and $k$ tests are generated using sampling with a temperature of 0.8.
\end{itemize}
For example, in a cascading system with three models, the parameter sets [$k_1=3$, $k_2=5$, $k_3=0$] and [$k_1=10$, $k_2=-1$, $k_3=3$] are valid. However, [$k_1=1$, $k_2=0$, $k_3=3$] is invalid because $k_2=0$ requires the system to exit at model 2, necessitating $k_3=-1$.

Similarly, for the number of test lines ($l$), we define $l \in \{0, 2, 4\}$ for each model. The value of $l$ is constrained such that $l=0$ if and only if $k=0$, indicating no test generation. For $l > 0$, the model generates a maximum of $l$ lines starting with \texttt{assert}. Note that the model may terminate generation before reaching $l$ lines, resulting in fewer than $k \times l$ test lines for the current model.

While $k$ and $l$ are model-specific parameters, $\theta$ is a global parameter shared across all cascading thresholds (e.g., from model 1 to model 2, and from model 2 to model 3). In our experiments, we sample $\theta$ from the set $\{0.0, 0.1, 0.2, \dots, 1.0\}$.

\paragraph{Sampling Validation Set}
To identify optimal parameters, we randomly sample a validation set $d_v$ from the full dataset $d$, reserving the remaining data for testing ($d_t$). In our experiments, $d_v$ comprises 30\% of the questions in $d$. 

\paragraph{Parameter Optimization}
On the validation set $d_v$, we execute the cascading system for all valid combinations of $[\theta, k_1, \dots, k_n, l_1, \dots, l_n]$, as described in Section \ref{subsect:cascading_pipeline}. For each combination, we compute the associated cost and accuracy. We then construct cost-accuracy plots for all $k$ and $l$ combinations at each value of $\theta$, analogous to the plots in Figure \ref{fig:all_test_pareto}. 

The optimal parameter combinations are identified as the \emph{Pareto points} in these plots---points for which no other combination achieves both higher accuracy and lower cost. These combinations are saved as the optimal parameters. For comparison, we also identify the optimal singular combinations by selecting Pareto points from scenarios where only one model is active (e.g., $k_1=-1$, $k_2=3$, $k_3=-1$). By comparing the Pareto points of cascading and singular combinations across plots, we determine the best value of $\theta$.

\paragraph{Evaluation}
Finally, we evaluate the selected cascading optimal combinations on the test set $d_t$ using the chosen $\theta$. We compare their performance against the singular optimal combinations to assess the effectiveness of the cascading approach. 

\begin{figure}[ht!]
    \label{fig:val_pareto}
    \centering
    \includegraphics[width=8cm, trim=0 0.9cm 0 0]{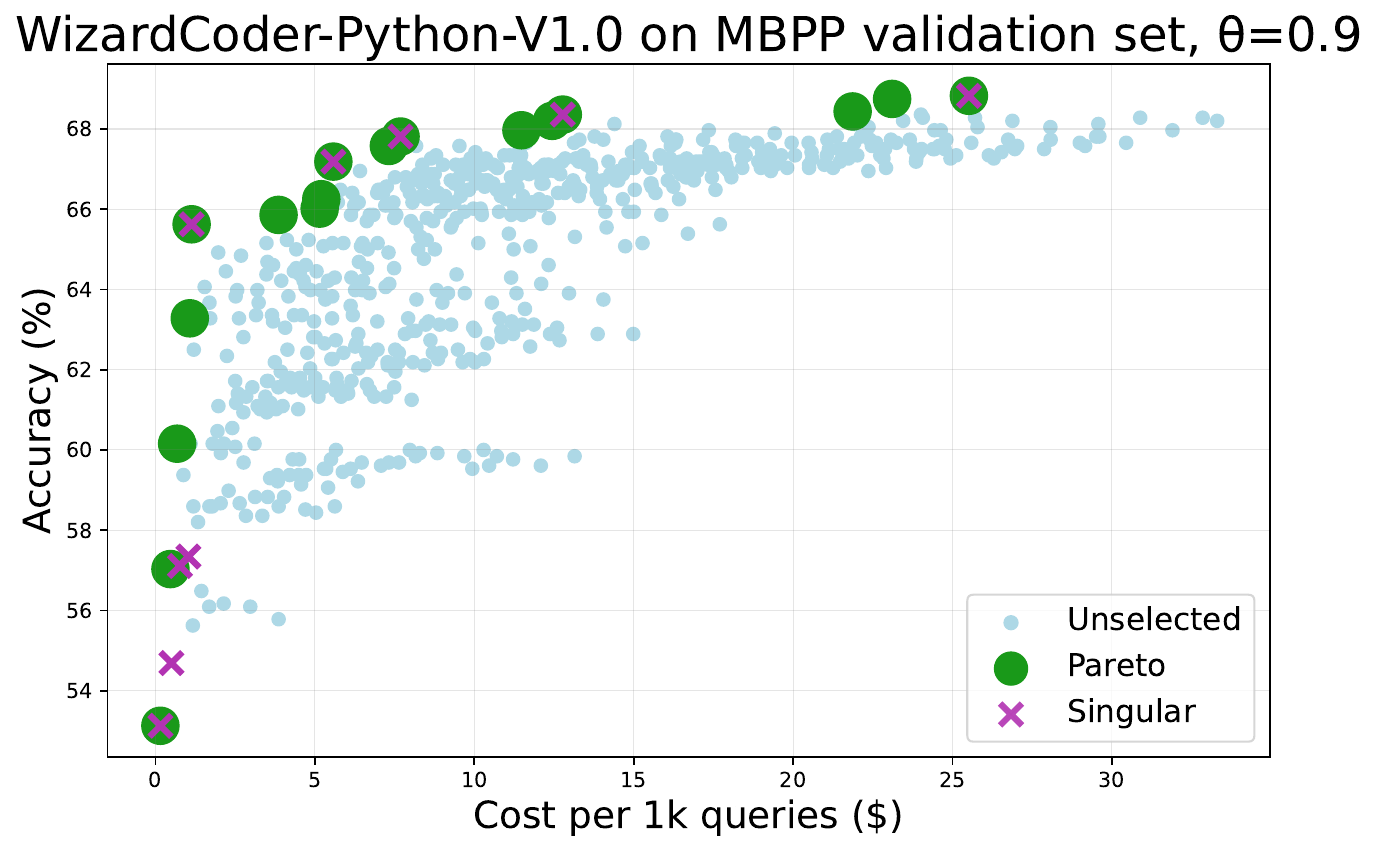}
    \vspace{0.1cm}
    \caption{Optimal parameter combinations selection by filtering the Pareto points in the cost-accuracy graph in validation set. The larger green dots are the selected Pareto points; the purple crosses are the Pareto points where only one model is used; the light blue dots are all the other plans with various combinations of models, numbers of answers and tests.}
\end{figure}
\vspace{-0.2cm}

\subsection{Cascading Threshold Parameter (\texorpdfstring{$\theta$}{theta})}
\label{subsect:theta}
\begin{figure}[hb]
    \centering
    \includegraphics[width=8cm, trim=2.5cm 1.2cm 2.5cm 1cm]{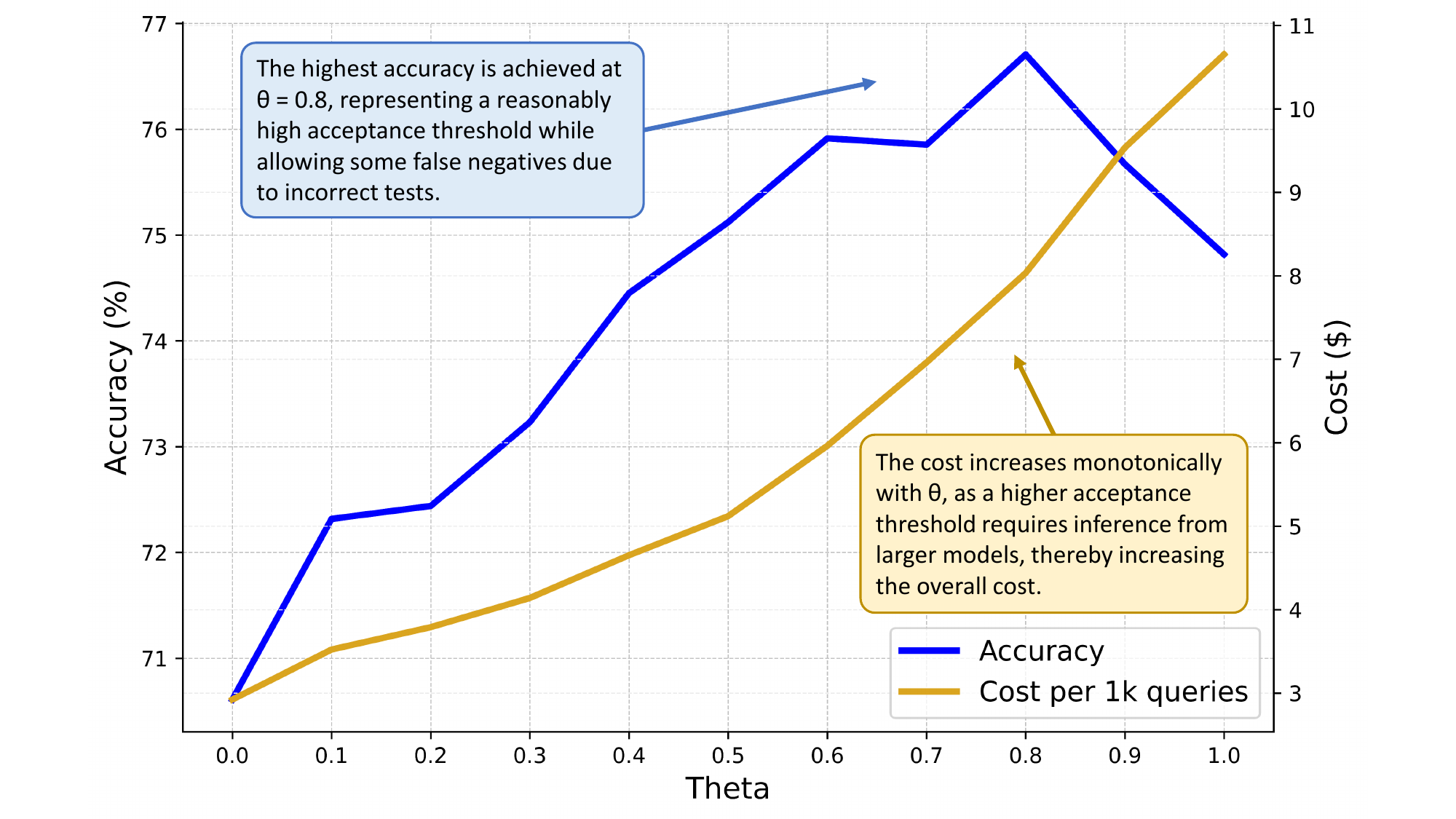}
    \vspace{0.2cm}
    \caption{Cascading result using the three models in Wizard-Python-V1.0 family on full HumanEval dataset, with $k_1=5$, $k_2=3$, $k_3=0$, $l_1=4$, $l_2=4$, $l_3=4$. We sample $\theta$ value on 0, 0.1, 0.2 \dots 1.0. The highest accuracy is 76.7\%, achieved at $\theta=0.8$ with a cost of \$8.03 per 1k queries.}
    \label{fig:theta}
\end{figure}

In this section, we analyze the impact of the cascading threshold parameter $\theta$ on the system's cost and accuracy. As described in Section \ref{subsect:cascading_pipeline}, $\theta$ determines when to escalate a query to the next larger model in the cascade. Specifically:
\begin{itemize}
    \item When $\theta$ is close to \textbf{0}, the system is more likely to accept the output of smaller models, as even modest solution quality meets the threshold. This results in lower computational cost but may compromise accuracy.
    \item When $\theta$ is close to \textbf{1}, the system becomes more stringent, frequently escalating queries to larger models. This increases computational cost but may improve accuracy, as larger models are more capable of generating high-quality solutions.
\end{itemize}

To illustrate the impact of $\theta$, we present an example in Figure \ref{fig:theta}. Using the WizardCoder-Python-V1.0 model family, we evaluated the full HumanEval dataset with fixed values for $k_1$--$k_3$ and $l_1$--$l_3$. Our results reveal two key trends:
\begin{itemize}
    \item \textbf{Cost}: As $\theta$ increases, the computational cost rises monotonically. This is because larger models are invoked more frequently, requiring additional computation.
    \item \textbf{Accuracy}: The highest accuracy is achieved at $\theta = 0.8$, rather than at $\theta = 1.0$. This counterintuitive result can be explained by the behavior of smaller models with larger $k$ values. When smaller models generate more solutions ($k$) and more tests ($l$), they have a broader pool of candidates to validate and select from. This increases the likelihood of identifying a correct solution, even if the model itself is less capable than larger models. Conversely, setting $\theta = 1.0$ imposes an overly rigorous standard, as it requires all solutions to pass all tests. This is highly unlikely for large $k$, as only a few solutions and tests may be correct. As a result, the system frequently escalates queries to larger models, missing potentially correct solutions from smaller models.
\end{itemize}

In summary, $\theta$ plays a critical role in balancing cost and accuracy. While higher values of $\theta$ favor the use of larger models, they do not consistently guarantee improved accuracy due to inherent trade-offs in solution validation and model constraints. In practice, servers typically operate within \emph{a specified budget range}. To address this, we determine the optimal value of $\theta$ by plotting the Pareto-optimal points from the validation set for each candidate value on a cost-accuracy plane. The optimal $\theta$ is then selected by identifying the curve that demonstrates the most Pareto-efficient behavior while offering the best cost coverage.

\subsection{Cost Calculation}
\label{sec:cost_calculation}
Since all models used in our experiments are publicly available, we estimate their computational cost based on inference time on GPUs. Our cost calculation mirrors the pricing schemes of commercial model providers like OpenAI, where the total cost is determined by multiplying the number of tokens generated by the per-token cost.

We deploy a server equipped with multiple NVIDIA GeForce RTX 3090 GPUs, each with 24GB of VRAM. For each model, we allocate $2^k$ GPUs, selecting the smallest $k$ that provides sufficient VRAM to run inference on a batch of 10 prompts. For each dataset, we randomly sample the maximum number of questions that can fit into a batch without exceeding the available VRAM. We then measure the total inference time $T$ and count the number of non-EOS tokens generated across all answers, denoted as $N_t$. 

To compute the per-token cost, we reference the hourly rental cost of an RTX 3090 GPU, $C = \$0.44/\text{hr}$ \cite{runpod_rent}. The per-token cost for each model is calculated as $c = T \times N_t \times C$. The cost statistics for all models on the HumanEval dataset are summarized in Table \ref{table:cost_stats_he}. Notably, we exclude the length of the input prompt from the cost calculation, even though it contributes to the computational cost. To address this, we collect averaged time statistics for each model family on each dataset, ensuring a more accurate representation of the overall cost.

\begin{table}[ht!]
\centering
\small
\newlength{\topspaceII}
\newlength{\bottomspaceII}
\setlength{\topspaceII}{0.25em}
\setlength{\bottomspaceII}{0.08em}
\begin{tabular}{>{\centering\arraybackslash}p{2.6cm} 
                >{\centering\arraybackslash}p{0.6cm} 
                >{\centering\arraybackslash}p{0.7cm} 
                >{\centering\arraybackslash}p{1.1cm} 
                >{\centering\arraybackslash}p{0.7cm} 
                >{\centering\arraybackslash}p{1.0cm} 
                }
\toprule
\addlinespace[0.32em]
\textbf{Model Family} & \textbf{Size} & \textbf{N.gpu} & \textbf{Time(h)} & \textbf{Batch} & \textbf{Cost(\$)} \\ 
\addlinespace[\bottomspaceII]
\midrule
\addlinespace[\topspaceII]
\multirow{3}{*}{\makecell{WizardCoder-\\Python-V1.0}} & 7B  & 2 & 2.17  & 20 & 1.91 \\ 
                                                     & 13B & 4 & 3.78 & 24 & 6.65 \\ 
                                                     & 34B & 8 & 5.75 & 48 & 20.24 \\ 
\addlinespace[\bottomspaceII]
\hline
\addlinespace[\topspaceII]
\multirow{3}{*}{\makecell{WizardCoder-\\V1.0}}       & 1B  & 1 & 0.31  & 180 & 0.13 \\ 
                                                     & 3B  & 1 & 0.58  & 80  & 0.26 \\ 
                                                     & 15B & 2 & 1.97  & 32  & 1.74 \\ 
\addlinespace[\bottomspaceII]
\hline
\addlinespace[\topspaceII]
\multirow{4}{*}{\makecell{Codegen-\\mono}}           & 350M & 1 & 2.36  & 48  & 1.04 \\ 
                                                     & 2B   & 1 & 5.97 & 12  & 2.63 \\ 
                                                     & 6B   & 2 & 6.44 & 12  & 5.67 \\ 
                                                     & 16B  & 4 & 9.00 & 12  & 15.84 \\ 
\bottomrule
\end{tabular}
\vspace{0.2cm}
\caption{Time and Cost stats per 1M tokens generated of all selected models on HumanEval dataset. We generated all completions with sampling. Time is averaged across 10 runs. }
\label{table:cost_stats_he}
\end{table}

The calculation is based on the gpus' \emph{full capacity} in running the LLMs to solve a \emph{maximum batch} of \emph{different} questions at the same time. The primary consideration is to even out the memory consumption of various model sizes and batch sizes. Furthermore, batching is by far one of the most effective ways for reducing energy cost \cite{10363447}, which is in accordance with our primary goal. 
For our experiments, we did not incorporate speculative decoding, because it has significantly decreased efficiency in large batching \cite{leviathan2023fast, qian-etal-2024-bass}. However, users could still incorporate speculative decoding when low-latency is strongly preferred. The target-draft model combination can be used to replace the single-model options, and it can generate multiple answers and tests in the same way as a single model. We leave more cost-efficient implementations of speculative decoding to future work.

\section{Experiment Setup}
\subsection{Model Families}
For our experiments, we selected three open-source code generation language model families: the Codegen-mono family (comprising four models with 350M, 2B, 6B, and 16B parameters), the WizardCoder-V1.0 family (comprising three models with 1B, 3B, and 15B parameters), and the WizardCoder-Python-V1.0 family (comprising three models with 7B, 13B, and 34B parameters). The Codegen-mono family models were trained on \textit{THEPILE}, \textit{BIGQUERY}, and \textit{BIGPYTHON} datasets. The latter two WizardCoder families were derived from the StarCoder and LLAMA-2 families \cite{touvron2023llama} and trained on the instructive dataset Code Alpaca \cite{codealpaca}. These models share the same architectures as their respective foundation models. 

The three model families have the following features that are ideal for our experiments: 
First, they are open-source and include models spanning a wide range of sizes. This diversity allows us to maximize inference time savings by adopting smaller models when their outputs meet the required quality threshold. 
Second, within each family, all models are trained on the same datasets, ensuring that larger models consistently achieve higher accuracy than smaller ones. If a larger model were to underperform compared to a smaller one, it would be eliminated from the cascading system, as it would not provide any benefit. 
Third, the three model families exhibit diverse performance. While newer models in Section \ref{chapter:related_work} achieve higher accuracy in code completion, we aim to demonstrate our method's robustness across all capability levels. Notably, each selected model was once the best in its size category.

\subsection{Datasets}
For our experiments, we utilize three datasets: HumanEval \cite{chen2021codex}, MBPP-sanitized \cite{austin2021programsynthesislargelanguage}, and the introductory-level subset of APPS-test \cite{hendrycks2021measuring}. These datasets contain 164, 427, and 1000 questions, respectively. The greedy accuracy of all models within each model family on these datasets is presented in Table \ref{table:model_dataset_cost_accuracy}. For the APPS-test dataset, we restrict our cascading results to the introductory-level questions using the two WizardCoder model families. This decision is motivated by the observation that the accuracy of other model families on introductory-level questions, as well as all model families on interview- and competition-level questions, falls below 10\% on average. When the test set is excessively challenging for a model family, the cascading system tends to waste computation on smaller models rather than saving computation, as the likelihood of smaller models producing acceptable solutions diminishes significantly.

In our experiments, for each run, we randomly sampled 30\% of the questions for the validation set, identified the Pareto-optimal combinations, and evaluated their cost and accuracy on the remaining 70\% of questions in the test set. To make sure that the validation set faithfully represents the difficulty of the test set, we checked that each model's accuracies in greedy search on both sets have a difference no bigger than 5\%.

\subsection{Environment}
Our server is equipped with NVIDIA GeForce RTX 3090 GPUs. We use Cuda Toolkit 11.8 and a PyTorch-based Python implementation from Hugging Face Transformers \cite{wolf-etal-2020-transformers} (version 4.31.0) with the Accelerate library \cite{accelerate}. This setup mirrors the training environment of WizardCoder. 
For each code completion, we generate a maximum of 1024 tokens per prompt with early stopping enabled.



\section{Results}
\label{chapter:results}
\begin{figure}[ht]
    \newlength{\hh}
    \newlength{\ww}
    \newlength{\moveback}
    \newlength{\legendhh}
    \setlength{\hh}{5.3cm}
    \setlength{\ww}{2.65cm}
    \setlength{\moveback}{-0.08cm}
    \setlength{\legendhh}{5.35cm}

    \hspace*{\moveback}
    \includegraphics[width=\hh, trim={\the\ww{} 0 \the\ww{} 0}, clip]{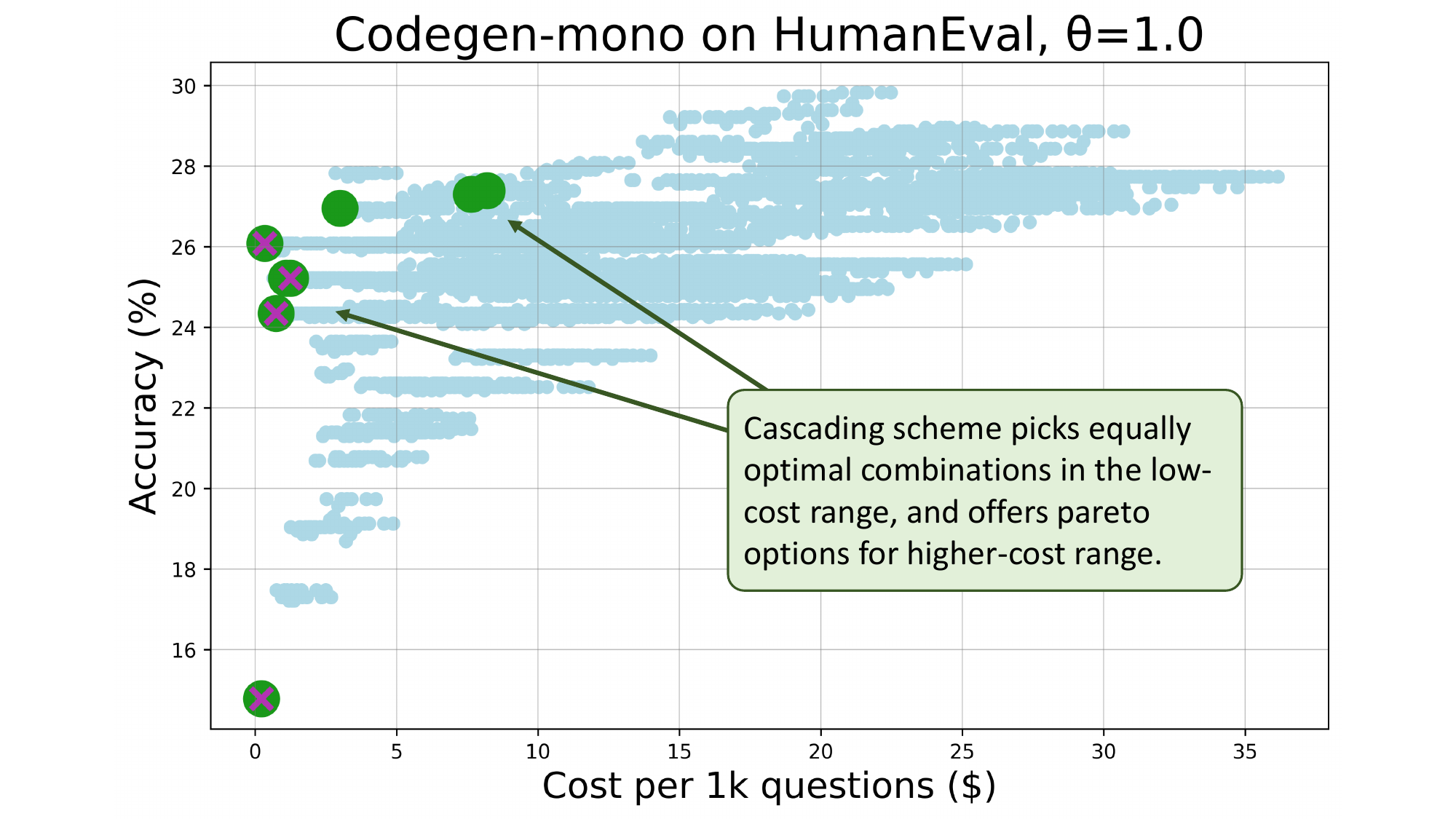} \
    \includegraphics[width=\hh, trim={\the\ww{} 0 \the\ww{} 0}, clip]{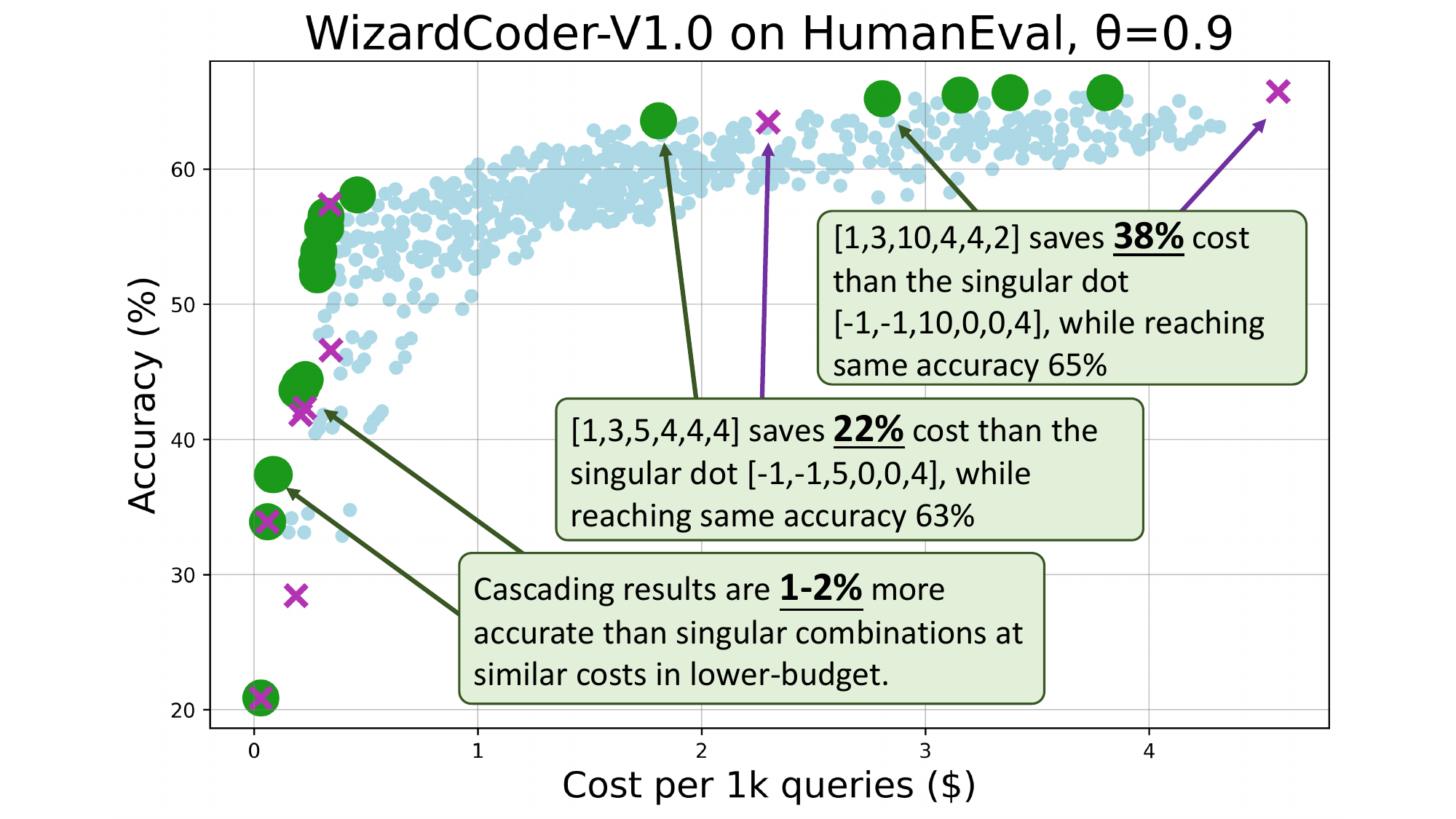} \
    \includegraphics[width=\hh, trim={\the\ww{} 0 \the\ww{} 0}, clip]{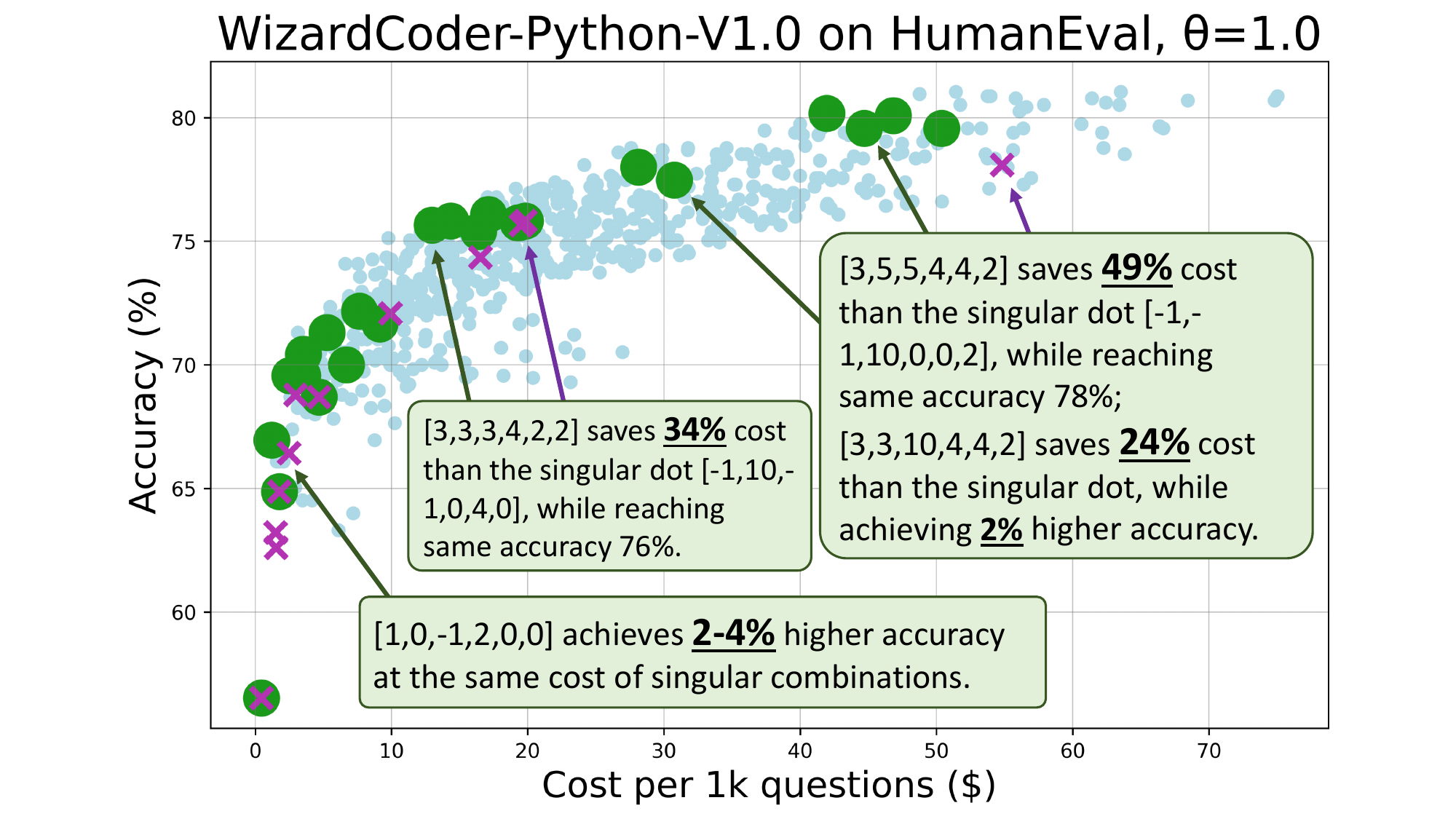}

    \vspace{0.2cm} 

    \hspace*{\moveback}
    \includegraphics[width=\hh, trim={\the\ww{} 0 \the\ww{} 0}, clip]{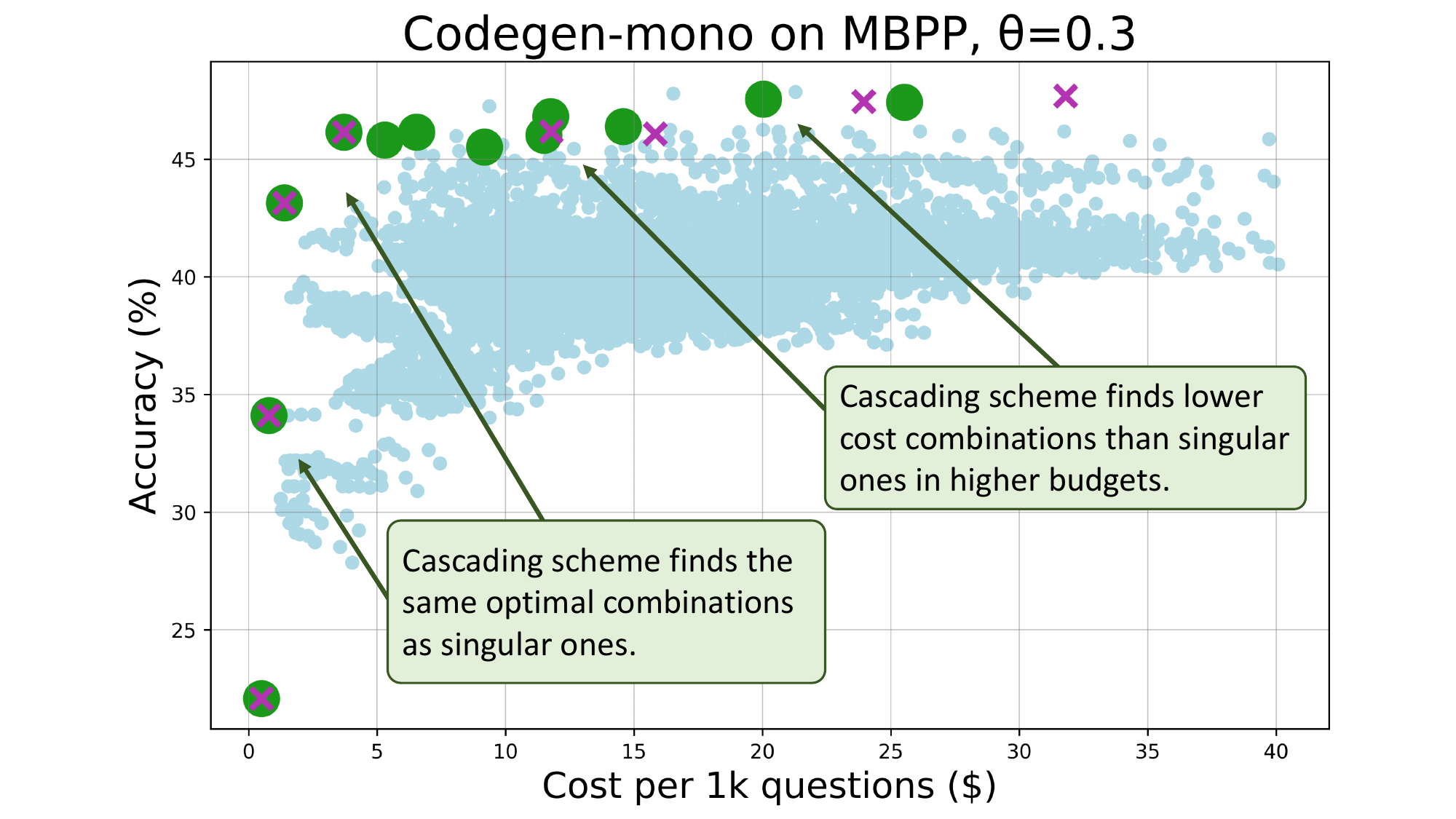} \
    \includegraphics[width=\hh, trim={\the\ww{} 0 \the\ww{} 0}, clip]{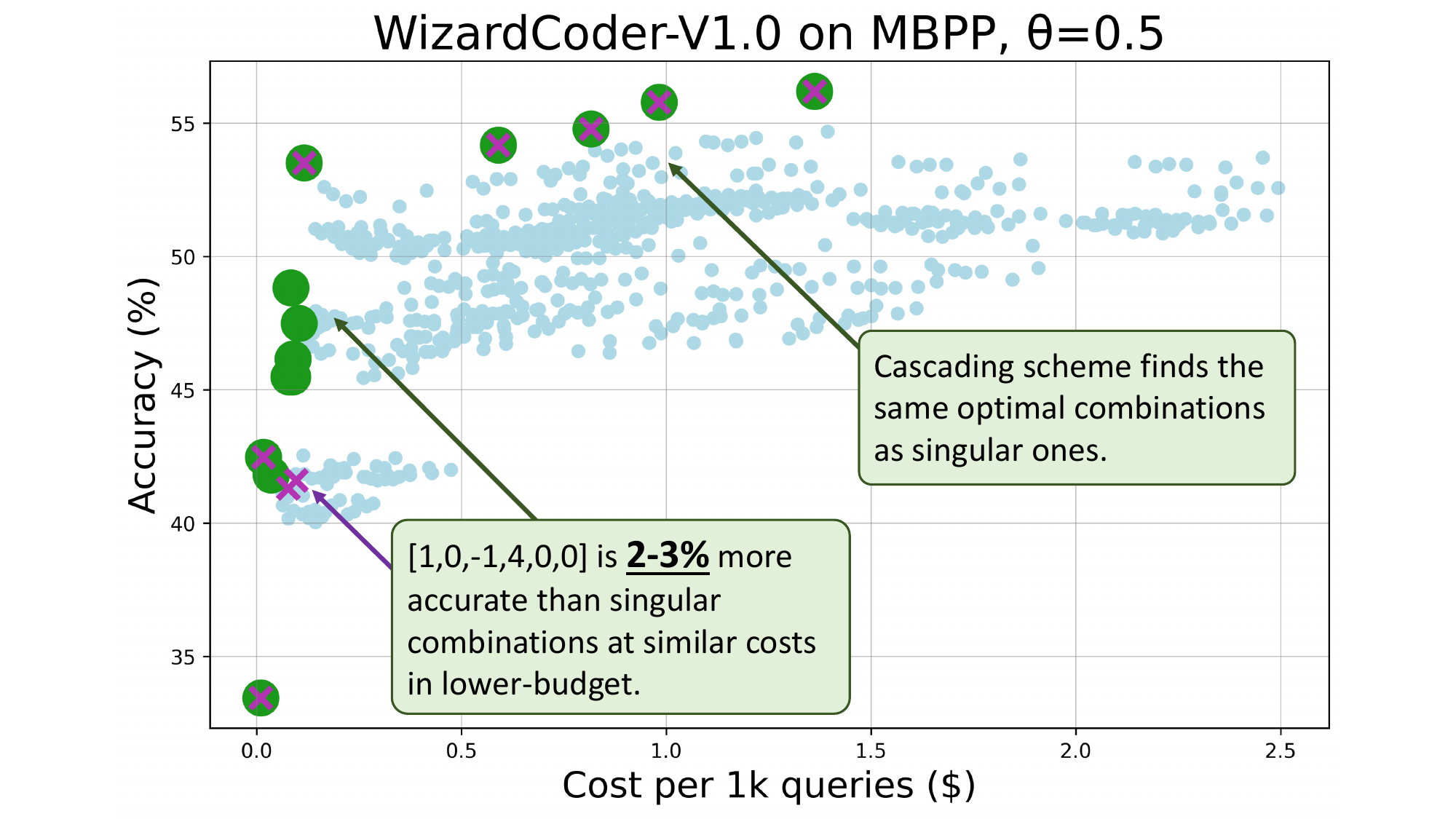} \
    \includegraphics[width=\hh, trim={\the\ww{} 0 \the\ww{} 0}, clip]{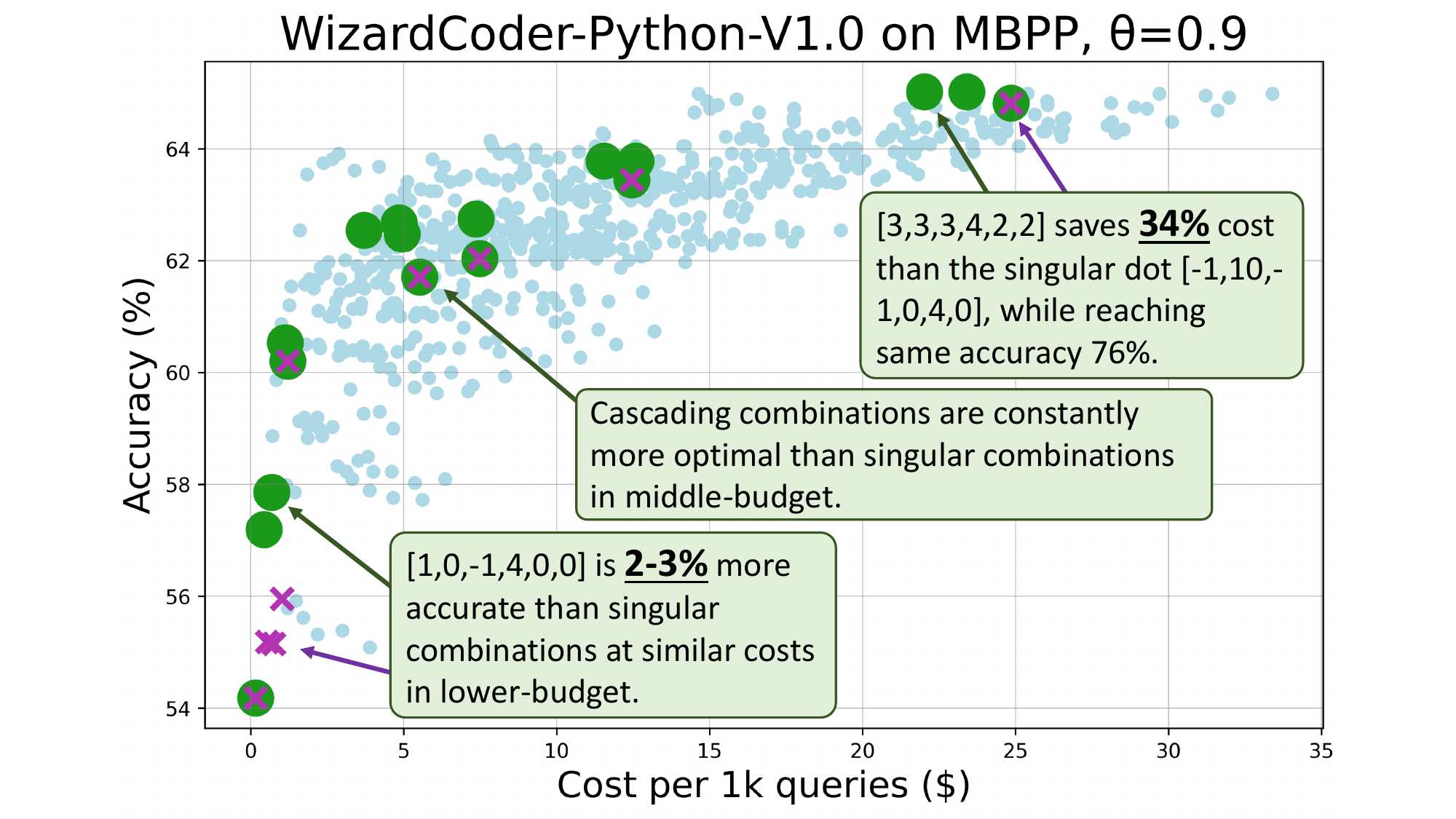}

    \vspace{0.2cm} 

    \hspace*{\moveback}
    \includegraphics[width=\legendhh, trim={0.40cm 0 0 0}, clip]{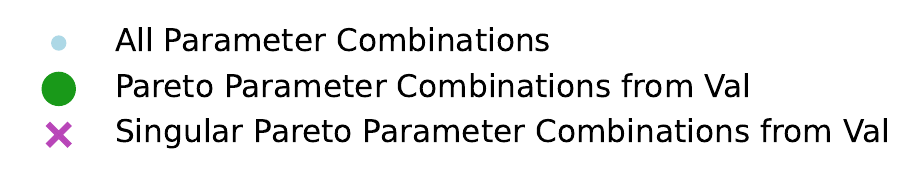} 
    \includegraphics[width=\hh, trim={\the\ww{} 0 \the\ww{} 0}, clip]{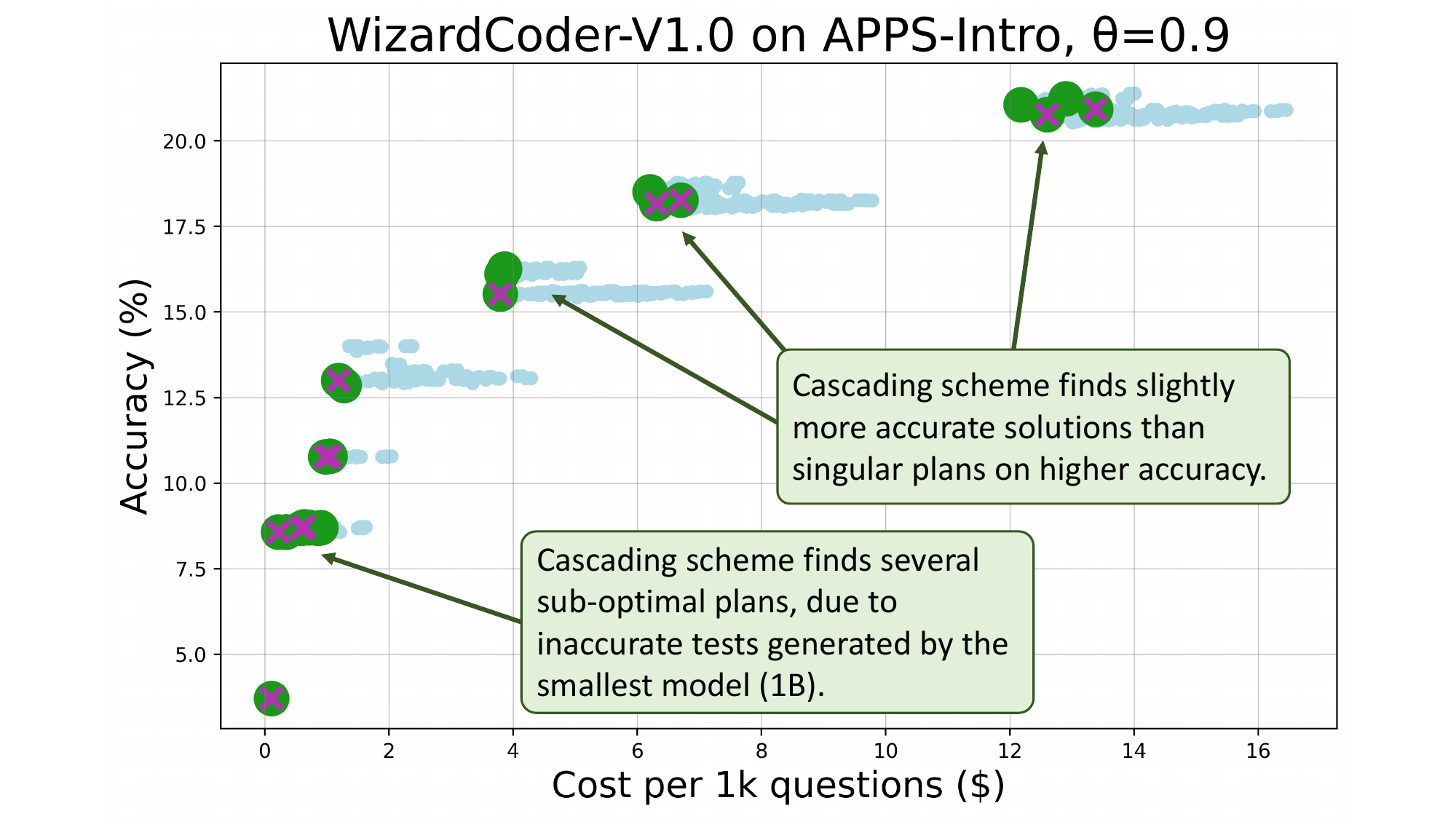} \
    \includegraphics[width=\hh, trim={\the\ww{} 0 \the\ww{} 0}, clip]{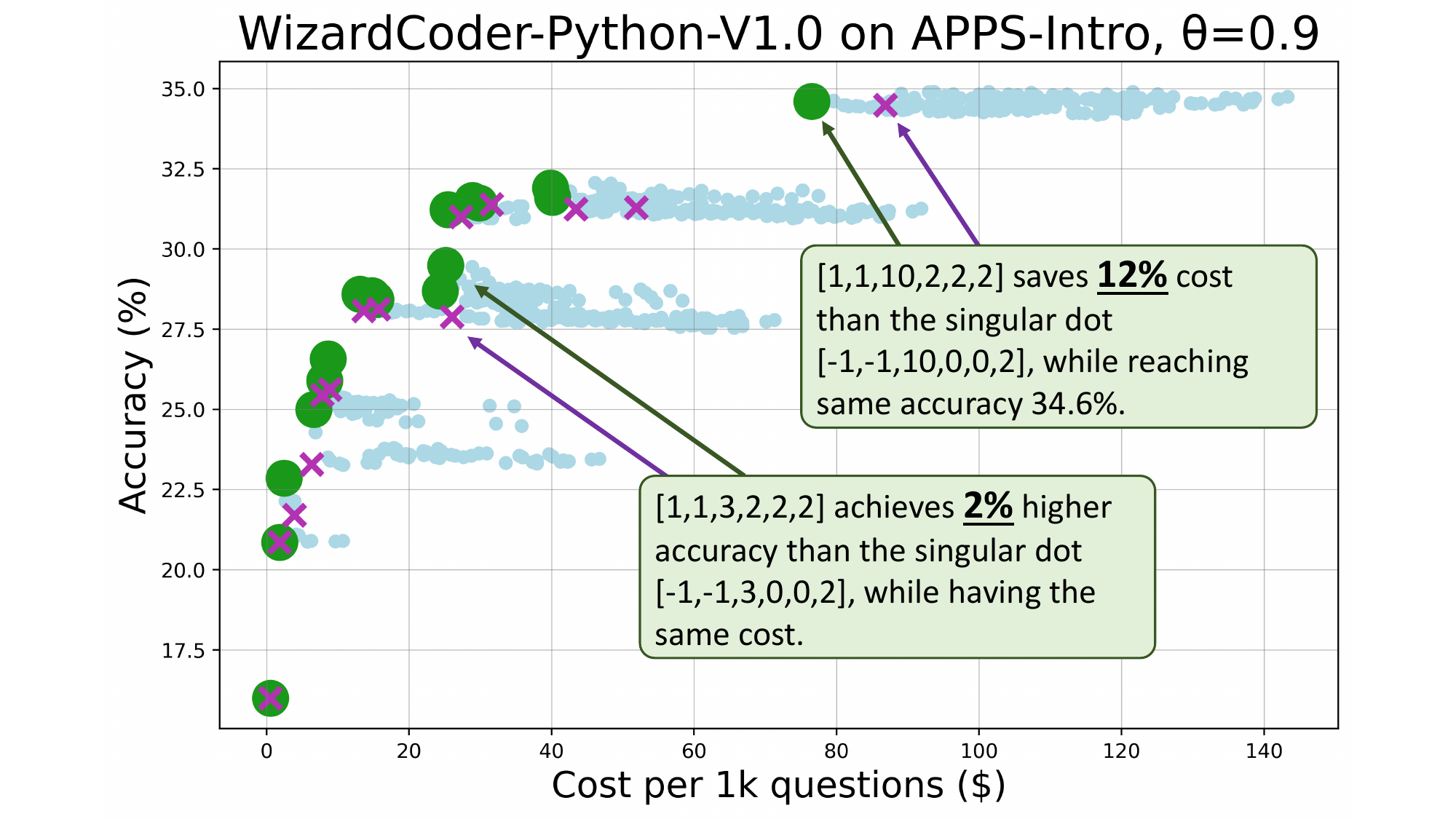}
    
    \caption{All $[k,t]$ combinations cost-accuracy results for Codegen-mono, WizardCoder-V1.0, WizardCoder-Python-V1.0 families on HumanEval, MBPP and APPS-Intro datasets, each plot with a selected optimal $\theta$. The two WizardCoder families have 3 models, and the Codegen-mono family has 4 models. Light blue dots represent all parameter combinations; green dots represent the cascading optimal combinations; purple crosses represent single-model optimal combinations. 
    All optimal combinations are selected from the validation set, following the procedure introduced in Section \ref{subsect:finding_parameters}.}
    \label{fig:all_test_pareto}
\end{figure}

Figure \ref{fig:all_test_pareto} illustrates the throughput and accuracy of the three model families on the HumanEval and MBPP datasets, as well as two model families on the APPS-Intro dataset. The green dots represent the Pareto-optimal parameter combinations selected from the validation sets using our method. 
In Table \ref{table:avg_cost_saving}, we present the average cost savings achieved by our scheme for each model family-dataset pair. Overall, our cascading scheme yields \emph{near-optimal} combinations across a \emph{wide budget range} for every model family and dataset.

\subsection{Comparison With Random Single-Model Self-Testing}
\label{sec:compare_avg}
We compare our selected self-testing plans against random single-model self-testing plans at the same accuracy level. The latter serves as a baseline, as prior works have focused solely on improving accuracy for individual models, neglecting cost efficiency and the utilization of other models within the same family. Without a careful validation-set selection, servers can only devise random self-testing plans with comparable accuracy. Notably, single-model performance is unaffected by $\theta$.

Figure \ref{fig:all_test_pareto} demonstrates that our cascading scheme consistently identifies near-Pareto solutions. To quantify the comparison, we calculate average cost savings across all accuracy ranges. For each cascading Pareto solution with cost $C_1$, we compute the average cost $C_0$ of all single-model solutions within the accuracy window $\pm 1\%$. The cost savings are then calculated as:
\[
\text{Cost Saving} = \frac{C_0 - C_1}{C_0}
\]
This process is repeated for all selected cascading Pareto solutions, and the overall average percentage savings is derived by averaging all calculated percentages. 

Our cascading scheme is most effective on the Codegen family, which includes four models compared to three in the other families. Additionally, the size ratio between the largest and smallest models is approximately 46x, significantly larger than the ratios for the other two families (5x and 15x, respectively), providing a broader range of cascading plans. However, from Table \ref{table:avg_cost_saving}, our scheme incurred the same level of costs for the WizardCoder-V1.0 family on APPS-Intro. The direct reason is that on the lower-accuracy range, our scheme finds several sub-optimal plans using the 1B and the 3B models. The root cause is that the 1B model has too low accuracy on the dataset (see Table \ref{table:model_dataset_cost_accuracy}). In such cases, it generates incorrect answers and tests, thereby increasing costs on wasted computations and decreasing accuracy on false positive tests. We there recommend excluding models with accuracy below 10\% when using the cascading scheme. Beyond this, there is no evidence linking question difficulty to average cost savings.

In Figure \ref{fig:avg_humaneval}, we apply Piecewise Cubic Hermite Interpolating Polynomial (PCHIP) \cite{fritsch1980monotone} to interpolate both our solutions and cost-averaged single-model solutions for each model family on HumanEval. The distance between the two curves highlights the cost savings achieved by our scheme.

\begin{table}[ht!]
\newlength{\topspaceIII}
\newlength{\bottomspaceIII}
\setlength{\topspaceIII}{0.25em}
\setlength{\bottomspaceIII}{0.06em}
\centering
\begin{tabular}{cccc}
    \toprule
    \addlinespace[\topspaceIII]
    Model Family & HumenEval & MBPP & APPS-Intro \\
    \midrule
    \addlinespace[\topspaceIII]
    WizardCoder-Python-V1.0 & 17.4\% & 30.8\% & 16.2\% \\
    \addlinespace[\topspaceIII]
    WizardCoder-V1.0 & 11.5\% & 11.4\% & -1.6\% \\
    \addlinespace[\topspaceIII]
    Codegen & 70.0\% & 39.5\% & - \\
    \bottomrule
\end{tabular}
\vspace{0.2cm}
\caption{Average cost savings of our model cascading scheme compared to random single-model self-testing scheme on same levels of accuracy.}
\label{table:avg_cost_saving}
\end{table}
\vspace{-0.4cm}

\subsection{Ablation Study: Comparison with Pareto Single-Model Self-Testing Solutions from Validation Set}
\label{sec:ablation_study}
To further validate the optimality of model cascading, we plot single-model Pareto solutions (purple crosses) on the same validation set in Figure \ref{fig:all_test_pareto}. In some cases, our Pareto solutions overlap with single-model solutions, as we consider scenarios where cascading is unnecessary. Nevertheless, our solutions consistently outperform single models in both accuracy and cost efficiency. The effect is particularly notable for the WizardCoder-Python-V1.0 family (right-most column), where cascading combinations yield superior results across datasets and budget ranges. This is attributed to the high capability of the models in this family, which generate high-quality tests even for challenging prompts. For the other families, although improvements over single models are less pronounced, our scheme accurately identifies single-model solutions when they are optimal.

\subsection{Cascading Threshold (\texorpdfstring{$\theta$}{theta})}
\label{sec:theta analysis}
The optimal $\theta$ values are typically high, either 1.0 (requiring answers to pass all tests) or 0.9. This reflects the high accuracy of generated tests, aligning with prior findings that LLMs generate better verifications than answers \cite{NEURIPS2023_91edff07, huang2024selfimprovementlanguagemodelssharpening}. However, as the number of test lines $t$ per test case increases, LLMs tend to make errors in the third or fourth lines, likely because input-output examples are provided in the prompts for HumanEval and APPS-Intro, allowing LLMs to copy values for initial test lines.

The value 0.9 also indicates tolerance for occasional false test lines. In the MBPP dataset, which lacks input-output examples, less capable model families exhibit lower optimal $\theta$ values due to higher probabilities of false negatives, reducing confidence in test results. Therefore, servers should select higher $\theta$ values when models are highly capable, and user instructions in prompts are more detailed.

\section{Conclusion}
We introduced a heuristic pipeline to identify near-optimal cascading strategies across a range of budgets, demonstrating its ability to achieve lower costs and higher accuracy for various model families and datasets. Our approach is fully black-box, making it compatible with any model family. While our experiments focused on Python code generation tasks, the method is language-agnostic and can be applied to other programming languages. 

In our experiments, we used a single global value for the cascading threshold parameter $\theta$. To further explore the potential of model cascading for code completion tasks, future work could investigate using different $\theta$ values for each model and each [$k, t$] combination. Additionally, integrating self-testing schemes with speculative decoding presents a promising direction for further research. Future works could also extend our method to incorporate speculative decoding for lower latency.

\vspace{1cm}
\bibliographystyle{unsrtnat}

\end{document}